\documentclass{elsart}
\usepackage{graphicx}
\usepackage{natbib}
\begin{document}
\begin{frontmatter}
\title{Techniques for measuring atmospheric aerosols at the High Resolution Fly's Eye experiment}

\author[Utah]{R.~Abbasi,}
\author[Utah]{T.~Abu-Zayyad,}
\author[LosAlamos]{J.F.~Amann,}
\author[Utah]{G.C.~Archbold,}
\author[Utah]{K.~Belov,}
\author[Columbia]{S.~BenZvi,}
\author[Montana]{J.W.~Belz,}
\author[Rutgers]{D.R.~Bergman,}
\author[Nevis]{J.~Boyer,}
\author[Utah]{C.T.~Cannon,}
\author[Utah]{Z.~Cao,}
\author[Columbia]{B.M.~Connolly,}
\author[Utah]{J.~Fedorova,}
\author[Columbia]{C.B.~Finley,}
\author[Utah]{J.H.V.~Girard,}
\author[Utah]{R.C.~Gray,}
\author[Rutgers]{W.P.~Hanlon,}
\author[LosAlamos]{C.M.~Hoffman,}
\author[Rutgers]{M.H.~Holzscheiter,}
\author[LosAlamos]{G.A.~Hughes,}
\author[Utah]{P.~H\"{u}ntemeyer,}
\author[Utah]{C.C.H.~Jui,}
\author[Utah]{K.~Kim,}
\author[Montana]{M.A.~Kirn,}
\author[Nevis]{B.~Knapp,}
\author[Utah]{E.C.~Loh,}
\author[Utah]{K.~Martens,}
\author[Tokyo]{N.~Manago,}
\author[Nevis]{E.J.~Mannel,}
\author[New Mexico]{J.A.J.~Matthews,}
\author[Utah]{J.N.~Matthews,}
\author[Utah]{J.R.~Mumford,}
\author[Columbia]{A.~O'Neill,}
\author[Utah]{R.~Riehle,}
\author[Utah]{K.~Reil,}
\author[New Mexico]{M.D.~Roberts*,}
\author[Columbia]{M.~Seman,}
\author[Rutgers]{S.R.~Schnetzer,}
\author[Utah]{P.~Shen,}
\author[LosAlamos]{G.~Sinnis}
\author[Utah]{J.D.~Smith,}
\author[Utah]{P.~Sokolsky,}
\author[Columbia]{C.~Song,}
\author[Utah]{R.W.~Springer,}
\author[Utah]{B.T.~Stokes,}
\author[Utah]{ S.B.~Thomas,}
\author[Rutgers]{G.B.~Thomson,}
\author[LosAlamos]{D.~Tupa,}
\author[Columbia]{S.~Westerhoff,}
\author[Utah]{L.R.~Wiencke**,}
\author[Rutgers]{A.~Zech}

\address[Utah] {High Energy Astroparticle Physics Institute, 
University of Utah, Salt Lake City, UT}
\address[New Mexico] {Department of Physics, University of New Mexico, Albuquerque, NM}
\address[Nevis] {Columbia University, Nevis Labs, Irvington, NY}
\address[Columbia]{Department of Physics, Columbia University, New York, NY}
\address[LosAlamos] {Los Alamos National Laboratory, Los Alamos, NM}
\address[Montana] {Department of Physics, Montana State University,Bozeman, MT}
\address[Rutgers]{Department of Physics and Astronomy, Rutgers University, Piscataway, NJ}
\address[Tokyo]{University of Tokyo, Institute for Cosmic Ray Research,
Kashiwa, Japan}

*Corresponding author email: roberts@cosmic.utah.edu

**Corresponding author email: wiencke@cosmic.utah.edu

\begin{abstract}

We describe several techniques developed by the High Resolution Fly's
Eye experiment for measuring aerosol vertical optical depth, aerosol
horizontal attenuation length, and aerosol phase function.  The
techniques are based on measurements of side-scattered light generated
by a steerable ultraviolet laser and collected by an optical detector
designed to measure fluorescence light from cosmic-ray air showers. We
also present a technique to cross-check the aerosol optical depth
measurement using air showers observed in stereo.  These methods can
be used by future air fluorescence experiments.
\end{abstract}

\end{frontmatter}

\section{Overview}
Cosmic-ray air fluorescence experiments measure light produced by high-energy 
air showers.  The technique is calorimetric, with the atmosphere
as the calorimeter.  The integrated scintillation light produced by an
extensive air shower is proportional to the the primary particle
energy and is almost independent of the primary particle composition.
However, once this light is produced, the amount that reaches the
observatory depends on how this light propagates from the shower
through the molecular and aerosol components of the atmosphere to the
light-collection optics of the detector.  The benefit of the
atmosphere that makes the air fluorescence technique possible is
balanced by the challenge to understand how the light propagates
through the atmosphere.

Several aspects of the High Resolution Fly's Eye (HiRes) experiment
address this challenge.  To reduce atmospheric effects, the experiment
is located in a remote desert where the average humidity and cloud
cover are relatively low.  The detector stations are located on hills
approximately 100 m above the desert floor.  This places the
optical detectors above many aerosols, including low-lying dust and
ground fog.  To measure atmospheric effects, the experiment includes
steerable uv lasers that probe the aperture of the
detector while the experiment is running.  The light scattered out of
a laser beam produces tracks in the same detectors that measure tracks
of uv light from air showers.  Changes in atmospheric aerosols change
the amount of light scattered and the amount of light that reaches the
detector.  Modeling the detector, the laser, and the atmosphere generates a
simulated set of detector measurements.  The simulated data are
compared and fit to the actual detector measurements of the laser.
This procedure yields a set of parameters that describe the aerosols.

\section{Components of the Atmosphere}
Most of the light observed from air showers is produced in the the
troposphere.  With regard to atmospheric fluorescence measurements,
the atmosphere has two main components: the molecular atmosphere and
aerosols.  Both components can be described in terms of optical
depths, denoted $\tau_{M}$ and $\tau_{A}$.  Assuming single
scattering, the transmission through a vertical column of atmosphere
can be expressed as $T=\exp(-\tau_{M}-\tau_{A})$.

\subsection{The Molecular Component}
The molecular atmosphere, comprising mainly nitrogen and oxygen
molecules, has absorption and scattering properties that are well
understood.  Because the 300 to 400 nm wavelengths of interest (see
\cite{K1996a} and \cite{N2004}) are much larger than the nitrogen and
oxygen molecules, the electric field can be approximated as constant
across a molecule, and consequently, light propagation can be
described using Rayleigh scattering theory [see, for example,
\cite{N2000a}, \cite{M2001}, and \cite{B1995a}].  The molecular
component is relatively stable, with small variations due to
height-dependent changes in temperature and pressure.  Because a
purely molecular atmosphere has the the clearest and best-understood
viewing conditions, it provides a boundary value solution to
light-propagation studies.  Air density can be obtained by applying
the ideal gas law to radiosonde measurements (collected at the Salt
Lake, UT, and Reno, NV, airports) of pressure and temperature as a
function of height.  The dependence on height follows an adiabatic
model, with variations at the 5\% level caused by temperature and
pressure changes associated with different seasons and weather
systems.  For circularily or unpolarized light, the scattering
probability as a function of scattering angle, $\theta$, follows
$1+cos^{2}(\theta)$, with higher-order corrections that are negligible
for the purposes of this work.

Applying the methods described in \cite{B1999} to Salt Lake radiosonde
data, yield values for $\tau_{M}$ at 355 nm, a wavelength that falls
near the middle of the air fluorescence spectrum.  $\tau_{M}$ from the
1.5 km elevation of HiRes (850 g/cm$^{2}$) through the entire atmosphere
is 0.50.  This value drops to 0.37 through 10 km (223 g/cm$^{2}$) above
ground, and to 0.17 through 3.6 km (557 g/cm$^{2}$) above ground, with
seasonal fluctuations of 0.005.  For comparison, we note the average
$\tau_{A}$ between ground and a 3.5 km height was measured at 355nm to
be 0.04 (see section 4).

\subsection{The Aerosol Component}
The aerosol component is more complicated. Aerosol sizes range from
10$^{-6}$ cm (large molecules) to 10$^{-3}$ cm (particulate matter).
The distribution of aerosols changes with location, height, and time.
Local dust can be lifted from the desert floor by wind and transported
to elevations typically less than 1 km.  More distant sources can
create aerosols that may be carried over large areas by upper-level
air flows.  A precise analytic description of absorption and
scattering is impossible to obtain because of the variable nature of
the aerosol size distributions and dielectric constants.  Light
scattering from aerosols of a specific size and spherical shape can be
described by Mie scattering theory (\cite{B1983}).  In general, the
scattering probability as a function of angle is characterized by a
forward peak, relatively little scattering at 90 degrees, and some
enhancement for backward scattering. However, because the precise
details of aerosol size, shape, and distribution are not known a
priori, one must rely on phenomenologically based models for
guidance. A commonly used simulation package, MODTRAN (Moderate
Resolution Transmission), was developed by the US Air Force
(\cite{L1988}).  The package contains aerosol models for different
regions including a ``US Standard Desert'' parameterization.

In contrast to the molecular atmosphere, aerosols are known to vary on
short time scales.  Weather fronts can load the atmosphere with
wind-driven aerosols. Rain and snowstorms can flush haze and dust from
the atmosphere, often leaving near-molecular viewing conditions.  The
aerosol density at ground level, the vertical distribution of the
aerosols, and the scattering and absorption properties can all change
within one night of observing.  Atmospheric monitoring at HiRes
therefore entails measuring the time dependence of the distribution of
aerosols in the detector aperture and determining their gross
properties to the level required to reconstruct air showers.

In this article, we describe the measurement of three aerosol
quantities: vertical aerosol optical depth ($\tau_{A}$), horizontal
aerosol scattering length at detector level (HAL), and aerosol phase
function.  $\tau_{A}$, measured from the elevation of the detector to
an elevation of 3.5km, is the most important quantity of the three for
several reasons.  First, because the field of view of a HiRes detector
begins at an elevation angle of 3 degrees, and because the HiRes
stations are on hills, none of the aperture extends to ground
level. At a distance of 19 km, the lower edge of the aperture is 1 km
above detector station height.  Second, $\tau_{A}$ describes the total
aerosol loading, and most of the aerosols lies below the height at
which most of the light is observed from air showers, especially for
showers above 10 EeV, which tend to be further away.  Therefore, under
the assumption of horizontal aerosol uniformity, (also addressed in
this article), most of the light reaching the detector is sensitive to
the total amount of aerosols; the correction for aerosol transmission
depends more on elevation angle than on distance.  As viewed along the
optical axis of a lowest ring HiRes detector, an uncertainty of 0.02
in ($\tau_{A}$) corresponds to an uncertainty of 11\% in transmission.
Over the full 30-3 degree range of elevation angles, this uncertainty
changes from 4\% at the very top to 32\% at the very bottom.  For
air-showers, these uncertainties translate to an uncertainty in energy
reconstruction of order 10 to 20\%.

\section{Apparatus}
Measurements of $\tau_{A}$ and HAL are determined from laser tracks
recorded by the HiRes detectors.

\subsection{The Steerable Laser System}
The steerable laser system, dubbed HR2SLS \cite{W1999},
fires a pattern of 1000-1300 shots per hour. Vertical shots can be
used to extract vertical aerosol optical depth. Horizontal attenuation
lengths can be extracted from near-horizontal shots fired across the
HiRes1 detector station.  A pattern of elevated shots fired in
15-degree steps in azimuth may be used to study atmospheric horizontal
uniformity.

The laser produces a 7 ns wide pulse of light at 355 nm.  This
wavelength closely matches the 357 nm fluorescence line. The maximum
beam energy is 7 mJ.  The beam can be steered in any direction above
the horizon.  The direction, energy, and time of each pulse are
measured and recorded locally.  The laser firing times are
synchronized using a GPS clock so that laser events can be isolated
from other HiRes triggers in a physics-blind manner. Different
energies and polarizations can be selected using filter wheels that
provide a combination of quarter wave-plates and attenuation filters.

One percent of the beam is sampled by a photo diode detector to obtain
a relative measurement of the pulse by pulse energy. Special
measurements are taken at regular intervals to determine the absolute
calibration.  For these measurements, an energy-absorbing pyroelectric
probe is placed temporarily where the beam exits to the sky.  The
pyroelectric probe is periodically recalibrated to 5\% by the
manufacturer using an NIST traceable system.

Circularly polarized beams with energies from 0.05 to 3 mJ were used
in obtaining measurements.  Circular polarization was chosen over
linear polarization to reduce polarization dependent scattering
effects and simplify simulations. The energy range must 
match the detector dynamic range for different beam directions.  For
example, vertical shots needed to measure $\tau_{A}$ are viewed at a
distance of 12.7 km.  These shots require energy of 1-3 mJ.  On the
other hand, the horizontal shots needed to measure HAL must pass
within a few hundred meters of the detector. These shots require a
much lower energy, typically 0.1 mJ, to avoid detector saturation.
For each combination of filter and direction setting, the laser fires
15 shots which are later averaged to reduce statistical uncertainties.

\subsection{The HiRes Detectors}
The detectors that view the laser tracks are described briefly.  HiRes
consists of two stations, HiRes1 and HiRes2, which are separated by
12.6 km.  The analysis described here used detectors (Abu-Zayyad et
al. 2000) at the HiRes1 station.  These detectors are arranged to
view nearly 360 degrees in azimuth from an elevation angle of 3-17
degrees.  A spherical mirror (3.78 m$^{2}$ effective area) at each
detector reflects light from the night sky through a uv pass filter
onto a cluster of 256 hexagonal photomultiplier tubes (PMTs).  Each
PMT views about 1x1$^{o}$ of night sky.  As a laser light pulse passes
upward through the field of view of a mirror, light scattered from the
beam is collected and focused to a spot that crosses downward across the cluster,
triggering PMTs along its path.  The nominal spot is about the size
of a PMT.  The nominal PMT gain is 10$^{5}$, and the nominal quantum
efficiency is 25\%.  The signal in a PMT from laser shots (and
air showers) is typically hundreds to thousands of photoelectrons.

The detector readout system at HiRes1 uses sample-and-hold
electronics.  Signal digitization and data formatting are performed by
local electronics at each mirror.  PMT outputs are AC coupled.  The
signal-processing channel used for this work has a three-pole Bessel
filter with a 375ns time constant and a charge integration gate of 5.6
$\mu$s.  TDCs measure times between PMT triggers by integrating a
gated constant-current source.  Discriminator thresholds are adjusted
every 4 seconds to maintain a count rate of 200 Hz.  The trigger
requirement for a signal digitization is a sixfold coincidence between
PMTs. Typical trigger rates are 5-10 Hz per mirror.  System dead-time
is less than 2\%.

\section {Determining Vertical Aerosol Optical Depth}
Under the assumptions of horizontal uniformity, and single scattering, 
the transmission, $T_{A}$, through an optical depth of aerosols, $\tau_{A}$,
at an elevation angle $\theta$ is
\begin{equation}
T_{A}=\exp(-\frac{\tau_{A}}{\sin(\theta)})  
\label{tau1}
\end{equation}

This equation includes both absorption and scattering.  For the case
that all the aerosols are between the starting and ending elevation of
interest, and the aerosols are uniform horizontally, $T_{A}$ is
independent of the aerosol vertical distribution.

\begin{figure}
 \vspace*{2.0mm} 
 \includegraphics[width=10.0cm]{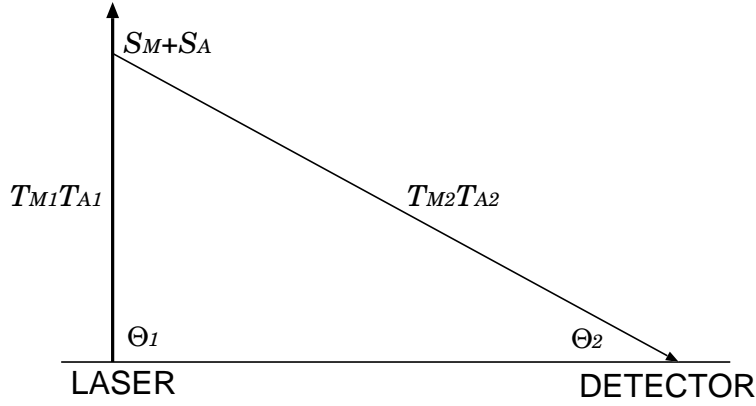} 
 \caption{Geometry for measuring aerosol optical depth.}
 \label{lasgeom}
\vspace*{5.0mm}
 \end{figure}

The method to determine $\tau_{A}$ normalizes the measurement of a
laser beam to the measurement that would be expected under
aerosol-free (i.e. molecular) atmospheric conditions.  Figure
\ref{lasgeom}  illustrates the geometrical configuration of the laser
and the detector.

The amount of light from the laser that reaches the detector can be
written as
\begin{equation}
N_{OBS}=N_{\gamma L}T_{M1}T_{A1}(S_A+S_M)T_{M2}T_{A2}
\end{equation}
where $N_{OBS}$ is the number of photons observed, $N_{\gamma L}$ is
the number of photons emitted per laser pulse, $T_{M1}$ and $T_{A1}$
are the molecular and aerosol transmission components from the laser
to the point of scatter, $S_{A}$ is the fraction of photons scattered
toward the detector by aerosols, $S_{M}$ is the fraction scattered by
the molecular component, and $T_{M2}$ and $T_{A2}$ are the molecular
and aerosol transmission components from the point of scatter to the
detector.  Normalized to the prediction for a purely molecular
atmosphere ($N_{MOL}=N_{\gamma_L}T_{M1}S_MT_{M2}$), equation 2 becomes
\begin{equation}
\frac{N_{OBS}}{N_{MOL}}=T_{A1}T_{A2}(1+\frac{S_A}{S_M})
\end{equation}
Applying equation \ref{tau1} yields 
\begin{equation}
\tau_{A}=\frac{-1}{\frac{1}{sin\theta_1}+\frac{1}{sin\theta_2}}(\ln(\frac{N_{OBS}}{N_{MOL}})-\ln(1+\frac{S_A}{S_M}))
\end{equation}
The two unknowns, $\tau_{A}$ and $S_{A}$, illustrate
the ambiguity between scattering and transmission inherent in
conventional LIDAR measurements.  The ambiguity can be eliminated if
 the scattering occurs in a region where $S_{A}$ is small compared to
$S_{M}$.  Equation 4 then simplifies to 
\begin{equation}
\tau_{A}=-\frac{sin\theta_1sin\theta_2}{sin\theta_1+sin\theta_2}\ln(\frac{N_{OBS}}{N_{MOL}})
\end{equation}

\begin{figure}
\includegraphics[width=13.0cm]{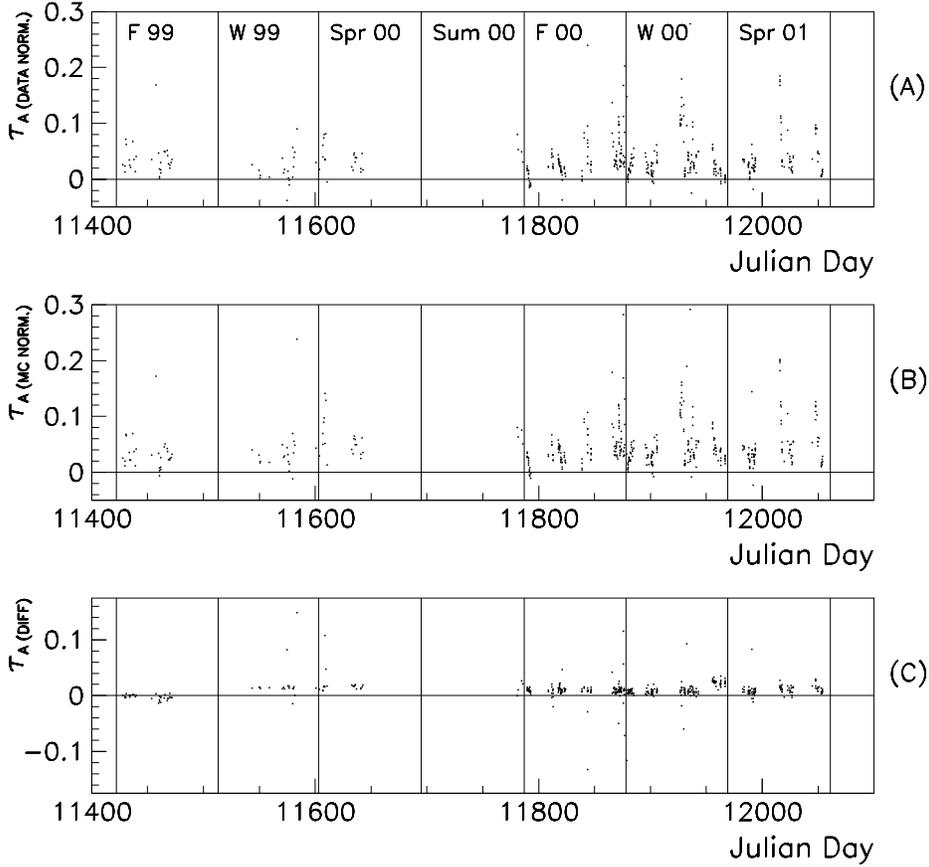}
\caption{Time evolution of $\tau_{A}$.  The top panel (A) shows
the data-normalized measurements.  The middle panel (B) shows the
Monte Carlo-normalized measurements, and the bottom panel (C) shows
the difference.}
\label{odtime}
\vspace*{5.0mm}
\end{figure}

\begin{figure}
 \vspace*{2.0mm} %
\includegraphics[width=13.0cm]{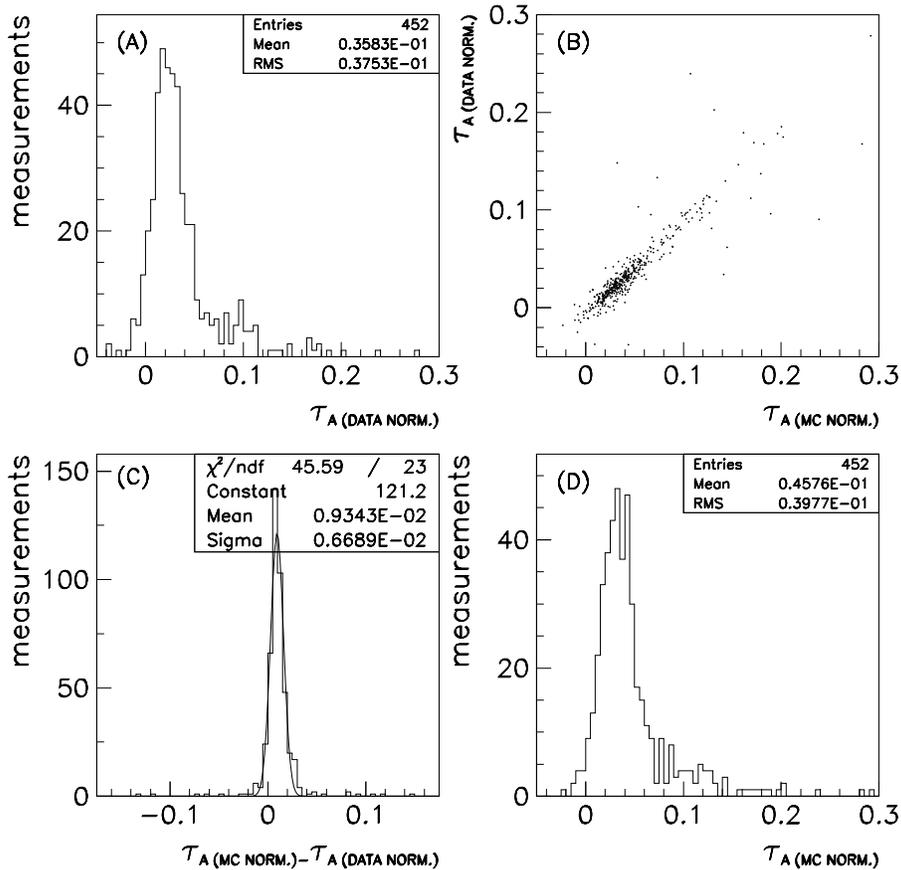}
        \caption{Vertical aerosol optical depth measured at 355 nm.  The upper left (A)
and lower right (D) panels show the measurement using the data-normalized 
and the Monte Carlo-normalized techniques. Panels (B) and (C) show the
correlation and the difference between the measurements.           
}
\vspace*{5.0mm}
\label{COMPnorm}
\end{figure}

In this case, $\tau_{A}$ becomes a simple function of geometry and the
number of observed photons that are normalized to $N_{MOL}$. $N_{MOL}$ can be
determined either by Monte Carlo simulation, or from data collected
under molecular conditions.  The latter method does not require an absolute
calibration of either the laser or the detector.  However, if the
reference data were collected during a period when aerosols were
present, the measured value of $\tau_{A}$ will be lower than the
actual value.  Normalizing to simulated data requires absolute
calibration of the laser and detector and simulation of light
propagation through a molecular atmosphere.  This analysis was
performed with data normalization and with Monte Carlo normalization.
The results are shown in figures \ref{odtime} and \ref{COMPnorm}.

The calculation of $N_{OBS}$ averages over sets of 15 vertical laser
shots fired each hour at two energy settings.  A region of PMTs was
selected at the top of the track.  For the data-normalized analysis,
the region was 2 PMTs wide by 4 PMTs high, corresponding to an
elevation of 3.5 km above the laser.  For the Monte Carlo-normalized
analysis, a 2x2 PMT grid was used, with a corresponding elevation of
3.8 km.

$N_{MOL}$ for the Monte Carlo normalized analysis included effects of
multiple-scattering, estimated to be at the 5\% level for a molecular
atmosphere.  To reduce computation time, the Monte Carlo simulation
exploited the symmetry of a vertical laser beam.  $N_{MOL}$ for the
data normalized analysis used data collected on December 12, 2000, UT hour 6.

The data sample in this analysis extends from September of 1999 to May 2001.  
Figure \ref{odtime} shows the time evolution of hourly $\tau_{A}$
measurements obtained by the two methods of analysis, and the
difference between the measurements.  The projection of these data onto the
$\tau_{A}$ axis and the correlation between them are shown in figure
\ref{COMPnorm}.  The systematic difference of 0.009 between the two
analyses could be caused by the presence of aerosols during the
reference ``clear period'' or by the relative calibration between the
laser and the detector.  For reference, 0.009 corresponds to a
calibration uncertainty of approximately 5\%.

The asymmetric shape of the $\tau_{A}$ distributions in figure
\ref{COMPnorm} are consistent with a molecular limit (at
$\tau_{A}$=0), and a tail on the high side due to aerosol loading.
The means of the $\tau_{A}$ distributions are 0.036 and 0.046.  Their
full widths at half maxima are 0.035.

We have identified three sources of systematic error that dominate the
measurement uncertainty.  A 10\% uncertainty in the relative
calibration between the laser and the detector contributes a
systematic error of 0.02 to the Monte Carlo-normalized $\tau_{A}$
average.  A 5\% time dependent variation in the gains of the PMTs used
in the analysis contribute 0.01 systematic error to both measurements.
Finally, the reference aerosol-free normalization data contributes an
uncertainty of 0.01 $\tau_{A}$ to the data-normalized measurement.  The
number of sets of laser shots analyzed reduces the statistical error of
the mean to 0.001.  Adding the systematic errors in quadrature yields

Mean $\tau_{A}$ = 0.046 $\pm$ 0.022$_{sys}$ $\pm$ 0.001$_{stat}$ (Monte Carlo-normalized)

Mean $\tau_{A}$ = 0.036 $\pm$ 0.014$_{sys}$ $\pm$ 0.001$_{stat}$ (data-normalized) 

We note that equation 1 yields a correction factor of 1.25 for light
traveling along the optical axis of a HiRes1 detector through a
$\tau_{A}$ of 0.04.

\begin{figure}
\includegraphics[width=13.0cm]{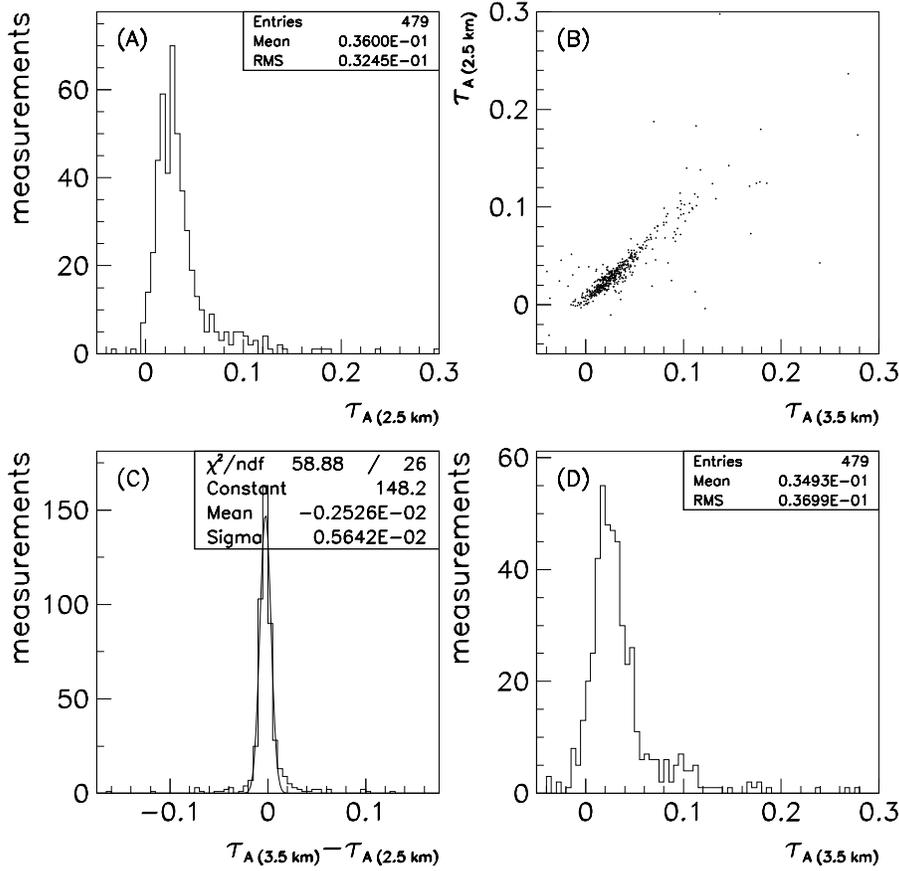}
\caption{Correlation (panel B) between aerosol optical depth as measured from 
z = 0 to 2.5 km (panel A) and from z=0 to 3.5 km (panel D). This
analysis uses data normalization.  If the average 0.04 $\tau_{A}$ was
uniformly distributed in height, an average difference of approximately
0.01 would have been observed.  This ``uniform height'' model is ruled out
because the difference (panel C) is much smaller.}
\vspace*{5.0mm}
\label{height-study}
\end{figure}

To study the vertical distribution of aerosols, the analysis was
repeated using sets of PMTs corresponding to different elevations.
The mean difference between $\tau_{A}$ for 0-3.5 km and 0-2.5 km was
found to be less than 0.0025. See figure \ref{height-study}. The mean
$\tau_{A}$ for 0-1.5 km was 0.008 smaller than mean $\tau_{A}$ for
0-3.5 km.  These numbers indicate that the bulk of the aerosols during
this period were distributed in the lower 1.5 km of the atmosphere.
They are also consistent with an average aerosol scale height of approximately
1 km.

\begin{figure}
\includegraphics[width=13.0cm]{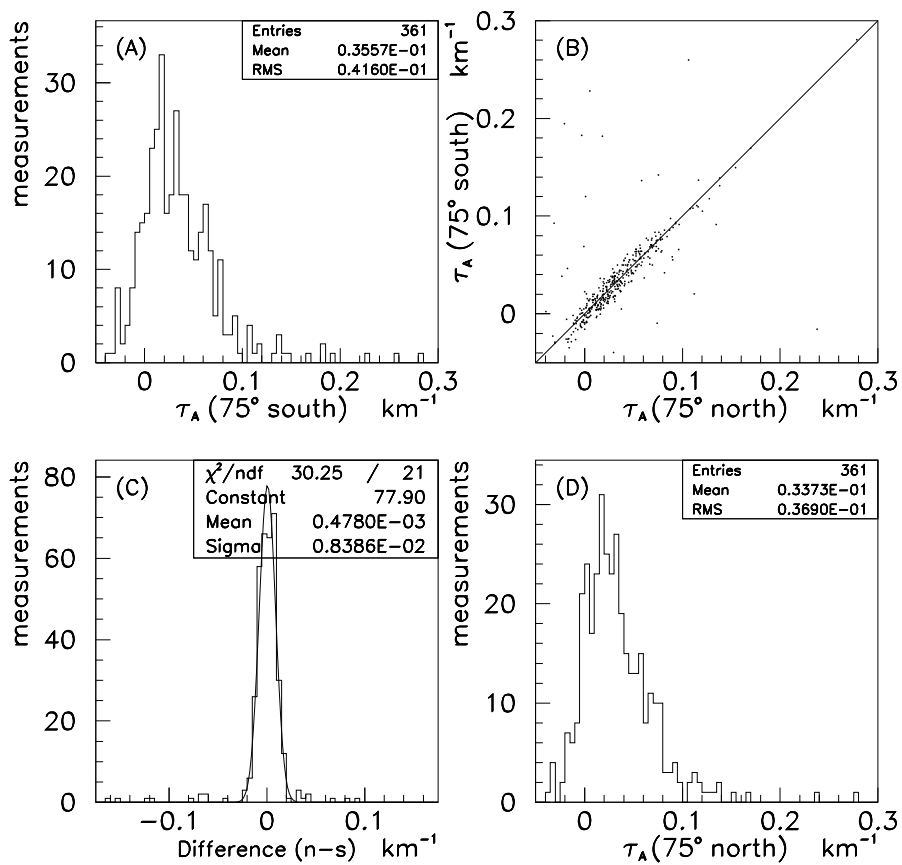}
\caption{Comparison of vertical aerosol optical depth from 15-degree
inclined laser shots fired at an azimuth angle of 75 degrees to the
north (panel A) and 75 degrees to the south (panel D) with respect to
HiRes1.  The regions of the tracks used for the measurements are
separated by roughly 20 km. The correlation (panel B) and difference
(panel C) provide a test of aerosol horizontal uniformity.}
\vspace*{5.0mm}
\label{horz-uniform}
\end{figure}

The optical-depth technique can also be applied to inclined laser
tracks.  To test horizontal uniformity, we compared optical-depth
measurements from two laser directions that were symmetric about the
line between the laser and the HiRes1 detector station.  The pattern
of laser shots includes shots of 15 degrees elevation angle and
azimuth angles of $\pm$ 75 degrees with respect to HiRes1.  The
portions of the tracks used in this measurement are 3 km above the
ground,and are separated by roughly 20 km.  Although the separation is
less than 20 km over most of the symmetric light paths (the light
starts at the laser and ends at the HiRes1 detector), the symmetric
measurements still probe different parts of the HiRes aperture and are
also made by different mirrors of the detector.  The left/right
differences in $\tau_{A}$ fall within $\pm$ 0.01 (see figure
\ref{horz-uniform}) for more than 90\% of the data.  A gaussian fit to
the difference in these $\tau_{A}$ measurements yields a sigma of
0.008.  In addition, the absolute scale of the average $\tau_{A}$, as
measured with inclined tracks, is consistent with the average obtained
from the analysis of vertical tracks.

\subsection{Horizontal Attenuation Length and Aerosol Phase Function}

\begin{figure}
 \vspace*{2.0mm}  
\includegraphics[width=10.0cm]{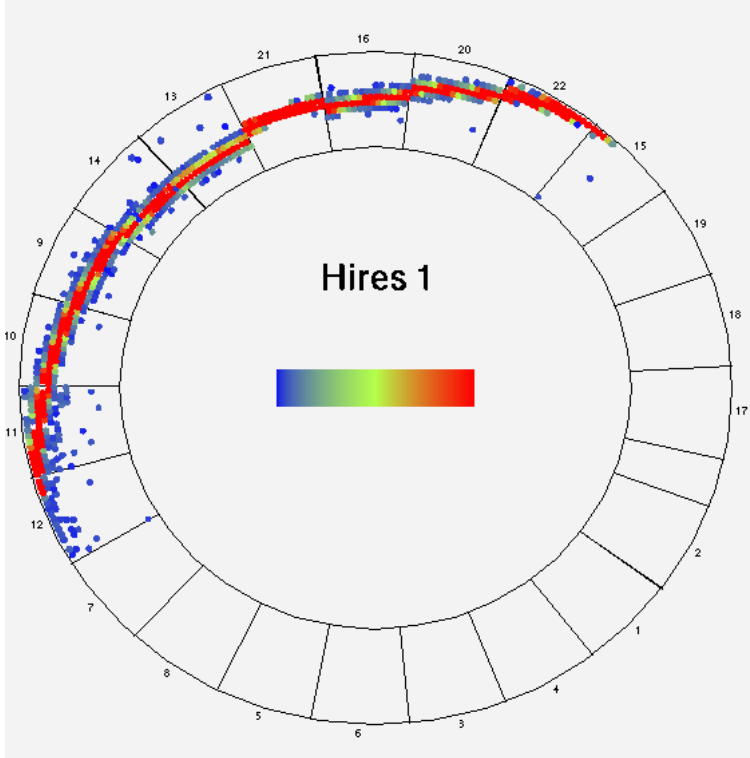}
\caption{Track made by nearly horizontal laser shot, as recorded by the 
HiRes1 detector.
}
\vspace*{5.0mm}
\label{h-track}
\end{figure}

\begin{figure}
 \vspace*{2.0mm}  
 \includegraphics[width=13.0cm]{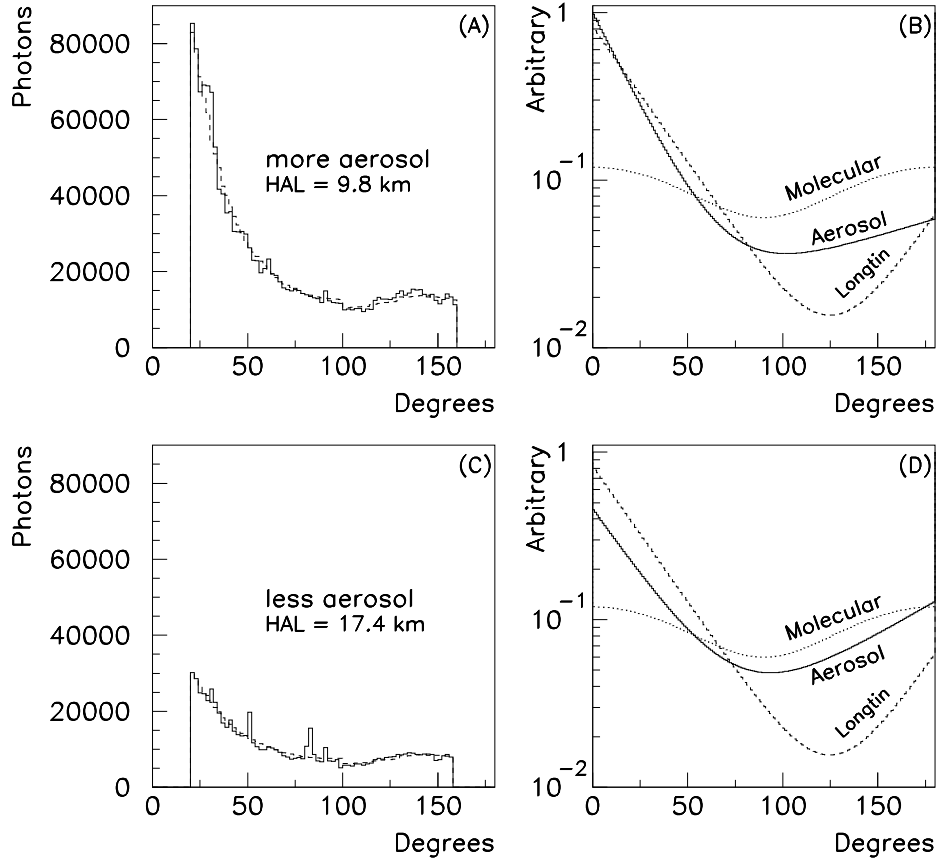} 

\caption {Two examples of fits for Horizontal Attenuation Length (HAL)
and aerosol phase function from data collected under two atmospheric
conditions.  Panels A and C show light intensity recorded from
near-horizontal laser shots at the HiRes1 detector as a function of
scattering angle.  Panels B and D show the corresponding aerosol phase
function obtained from the fit.  For comparison, the molecular phase
function and the Mie theory-derived phase function of the Longtin
desert aerosol model are also shown.}
 \vspace*{5.0mm}
 \label{PF}
 \end{figure}

The horizontal attenuation length (HAL) is the aerosol attenuation
length at ground level. HAL provides a normalization of the aerosol
density.  The phase function, or normalized differential
scattering cross section, of the aerosols is needed to correct for
Cherenkov light scattered from an air-shower.  HAL and phase function
are measured simultaneously with a fit to a near horizontal shot from
the HR2SLS.  This shot, passing within 450 m of the HiRes1 site,
generates a long track in the detector (see figure \ref{h-track}) that
is observed over a large range of scattering angles.  Assuming
horizontal uniformity, an aerosol scattering albedo near unity, and
known molecular scattering conditions, the longitudinal track profile
can be fit to determine HAL and the aerosol phase function (see figure
\ref{PF}).

The functional form of the aerosol phase function is

\begin{equation}
\frac{d\sigma}{d\Omega} \propto e^{-{\rm B}\theta}+{\rm C}e^{{\rm D}\theta}
\label{mie_pf}
\end{equation}

where $\theta$ is the scattering angle in radians and ${B, C, D}$ are free
parameters in the fit. This functional form reproduces the shape of a Mie 
theory-derived phase function except in the very forward
scattering region ($<2^{\circ}$), where only a small fraction of the
total scattered signal appears.

\begin{figure}
\includegraphics[width=13.0cm] {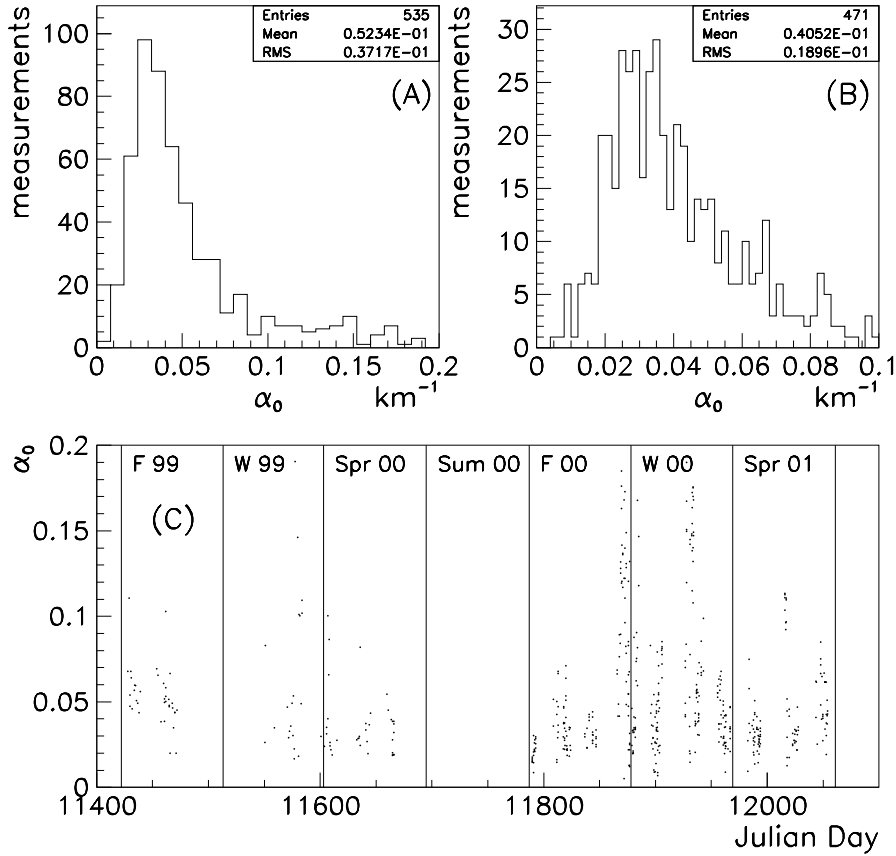}
\caption{Measurements of $\alpha_0$ from near-horizontal laser
shots.  $\alpha_0$ is the inverse of the horizontal aerosol scattering
length at the height of the detector.  Panels A and B show the
distribution on different scales.  The time dependence is shown in
panel C.
}
\vspace*{5.0mm}
\label{xa}
\end{figure}

Figure \ref{xa} shows the measured distribution of ground-level
scattering coefficients ($\alpha_0$ = HAL$^{-1}$).  If the very hazy
data ($\alpha_0>0.1$ km$^{-1}$) are excluded, the average value of
$\alpha_0$ is 0.04 km$^{-1}$.  

A mean aerosol scale height can be inferred by dividing the mean
$\tau_{A}$ by the mean $\alpha_0$. This calculation yields 1 km.  We
note this value is consistent with the differences between the
$\tau_{A}$ measurements at different heights as described previously.

\section{Checking the Aerosol Optical Depth using Air-Shower Data}

A sample of air-showers observed by both HiRes1 and HiRes2 was used to
cross-check the scale of the atmospheric laser measurements.  The
technique, detailed elsewhere (\cite{A2005}), is based on the
principle that multiple eyes viewing a common track segment should
reconstruct the same number of emitted photons after corrections for
geometrical factors and atmospheric attenuation.  If two detectors
view the same shower through different optical depths, an error in the
atmospheric model can cause an inconsistency between the two detectors
in the number of emitted photons from the common track segment.
Although this technique is a useful cross-check, the detection rate of
air showers is too low to provide precise measurements over a single
night.

For this study, a sample of 415 stereo events recorded between August
25, 2000, and May 23, 2001, was selected.  To ensure well-constrained
stereo reconstruction, the opening angle between the shower-detector
planes was required to be between 8 and 172 degrees. A cut on the
minimum viewing angle between the shower axis and the mirror axes of
30 degrees at each detector reduced asymmetric contributions from
scattered Cherenkov light. For each event the light-balance $\Delta_N$
was calculated from the number of photons reconstructed by the HiRes1
and HiRes2 detectors from the common track segment,
\begin{equation}
\Delta_N \equiv
ln \left( \frac{ N_{HiRes1} }{ N_{HiRes2} } \right)
\label{lightbal}
\end{equation}
Light-balance is plotted as a function of the linear asymmetry $\Delta_r$,
\begin{equation}
\Delta_r \equiv
r_{HiRes2} - r_{HiRes1}
\end{equation}
where $r_{HiResX}$ is the distance between a detector and the center of the
common segment.

\begin{figure}
 \vspace*{2.0mm}
 \includegraphics[width=14.0cm]{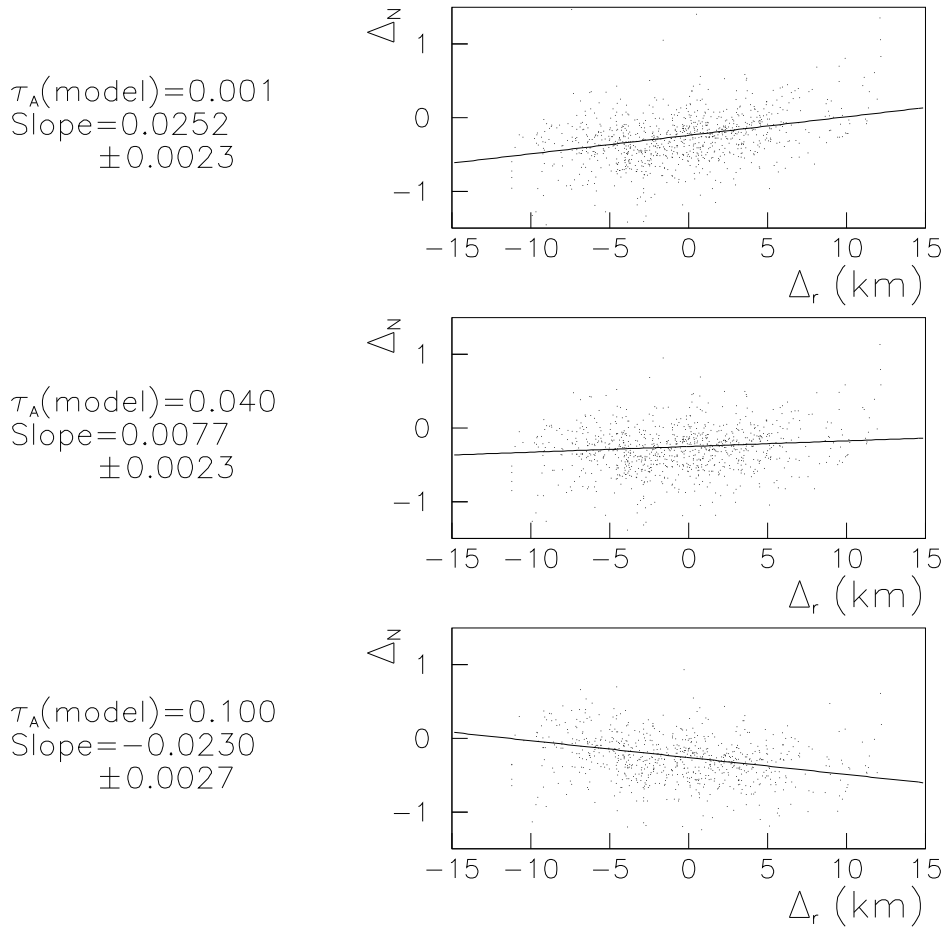}
 \caption{Light balance as determined by applying three different
aerosol models to a single set of stereo events:
top panel: near molecular, $\tau_{A}$=0.001; 
middle panel: $\tau_{A}$ = 0.040 
bottom panel: $\tau_{A}$ = 0.100; 
(see text).
}
 \label{LBal}
\vspace*{5.0mm}
 \end{figure}

Figure \ref{LBal} shows this analysis for three aerosol models: 1)
near molecular $\tau_{A}$ = 0.001; 2) $\tau_{A}$ = 0.040 HAL = 25; 3)
``Standard Desert'' $\tau_{A}$ = 0.100 HAL = 12 km. The Standard
Desert parametrization leads to an over correction.  On average, the
far detector reconstructs a ``brighter'' common segment than does the
near detector.  Conversely, if one assumes a complete absence of
aerosols (i.e. molecular atmosphere), a systematic under correction is
observed. The 0.04 $\tau_{A}$ 25 km HAL model, obtained from laser
measurements, falls between these extremes, yielding the smallest
slope of the three.  A $\tau_{A}$ of 0.055 yields a flat slope.  The
systematic error in this method is approximately 0.01.

\section{Conclusions}
We have developed several methods to measure the atmospheric aerosol
properties of optical depth, extinction length, and phase function.
These methods use side-scattered light from a 355 nm uv laser that
produces tracks in the same optical fluorescence detectors that
measure tracks from extensive air showers.  We have measured
distributions of aerosol optical depth and horizontal extinction
length at the location of the High Resolution Fly's Eye experiment.
We found the distribution of aerosols is horizontally uniform over a
distance of 20 km to less than 0.01 in $\tau_{A}$ on more than 90\%
of the nights observed, and that most of these aerosols are in the
lower 2 km of the atmosphere.  Between fall 1999 and spring 2001, the
optical depth of the aerosol component is on average less than 25\% of
the optical depth of the molecular component.  As a cross check, we
note that the average $\tau_{A}$, as obtained by laser measurements,
is consistent an average determined by comparing commonly viewed
segments of extensive air showers observed by the two HiRes
fluorescence detectors.

\section{Acknowledgments}

This work is supported by US NSF grants PHY-9321949, PHY-9322298,
PHY-0998826, PHY-024528, PHY-0305516, PHY-0307098, and by the DOE
grant FG03-92ER40732.  We gratefully acknowledge the contributions of
the technical staffs at our home institutions.  The cooperation of
Colonels E. Fischer and G. Hatter, the US Army, and Dugway Proving
Ground staff is greatly appreciated.

\end{document}